\documentclass[preprint,aps,floatfix]{revtex4-1}

\usepackage{graphicx}
\usepackage{graphics}
\usepackage{color}
\usepackage{subfigure}
\usepackage{amsmath}
\usepackage{amssymb, amsthm}
\usepackage{txfonts}

\begin{document}

\title{Truncated Wigner Method for Bose Gases}
\label{ruostekoski}

\author{J. Ruostekoski and A.D. Martin}

\address{School of Mathematics, University of Southampton,
Southampton, SO17 1BJ, United Kingdom}

\begin{abstract}
We discuss stochastic phase-space methods within the truncated Wigner approximation and show explicitly that they
can be used to solve non-equilibrium dynamics of bosonic atoms in one-dimensional traps. We consider systems
both with and without an optical lattice, and address different approximations in the stochastic
synthesization of quantum statistical correlations of the initial atomic field.
We also present
a numerically efficient projection method for analyzing correlation functions of the simulation results, and demonstrate physical examples of non-equilibrium quantum dynamics of solitons and atom number squeezing in optical lattices.
\end{abstract}

\maketitle

\section{Introduction}

In stochastic phase-space methods based on sampling classical probability distributions the common
approach is to unravel the evolution dynamics into stochastic trajectories each of which obey the classical
mean-field dynamics, with or without additional dissipative coupling terms to environment.
Each trajectory is a representative of a probabilistic initial state distribution that is numerically generated by a Monte Carlo type of sampling. The probability distribution is selected in order to synthesize as closely as necessary, e.g., the thermal distribution or quantum statistical correlations of the initial state. The phase-space representation that most accurately reproduces classical mean-field dynamics is the Wigner representation because of the `correct amount' of quantum noise in the initial state \cite{gardiner_zoller_book_99,sinatra_lobo_02}. It has become common to call mean-field dynamical simulations together with sampling of quantum noise as the Truncated Wigner Approximation (TWA) \cite{drummond_hardman_93,steel_olsen_98,sinatra_lobo_02,isella_ruostekoski_05,isella_ruostekoski_06,blakie_bradley_08,polkovnikov_10,martin_ruostekoski_10a}.

In this Chapter, we consider the unitary evolution in the TWA formalism in ultra-cold atom systems without additional dissipative coupling to an environment (for TWA employed in an open system, see for instance Ref.~\cite{javanainen_ruostekoski_11}) that is particularly suitable for analyzing dissipative non-equilibrium quantum dynamics in 1d systems. The TWA approach is able to represent systems with a large number of degrees of freedom using a stochastic representation of the atomic field operator. In this method, dissipative dynamics emerge from a microscopic treatment of the unitary quantum evolution, due to energy dissipation within the large phase-space without any additional explicit damping terms in the Hamiltonian. Quantum and thermal fluctuations of the atoms are included in the initial state, and the resulting quantum statistical correlations of the initial state may be accurately synthesized for different quantum states in the Wigner representation.

Atomic systems with enhanced quantum fluctuations that may be modeled with TWA can, for instance, be prepared in tightly-confined cigar-shaped atom traps, where the strong transverse confinement suppresses density fluctuations along the radial direction of the trap (see, e.g., Ref.~\cite{tolra_ohara_04,kinoshita_wenger_06,fertig_ohara_05,mun_medley_07}). Quantum effects may be further strengthened by reducing the kinetic energy of the atoms by means of applying an optical lattice potential along the axial direction \cite{fertig_ohara_05,mun_medley_07}.

In this Chapter we briefly summarise the TWA method, before addressing different approximations in synthesization of quantum and thermal noise in the initial state. We start with a simple uniform system and phonon excitations within the Bogoliubov approximation. These are extended to non-uniform systems and situations where the back-action of the excited-state correlations on the ground-state atoms is included in a self-consistent manner. A particular problem of analyzing non-equilibrium quantum dynamics in TWA, related to symmetric operator-ordering of the Wigner distributions is addressed by providing a numerically practical projection technique.

Applications of the approaches presented here include the studies of dissipative non-equilibrium systems both with and without an optical lattice, such as the fragmentation of a BEC by ramping on an optical lattice \cite{isella_ruostekoski_05,isella_ruostekoski_06,gross_esteve_11}, dissipative atom transport \cite{ruostekoski_isella_05}, dynamically unstable lattice dynamics \cite{shrestha_javanainen_09}, and dark solitons \cite{martin_ruostekoski_10a,martin_ruostekoski_10b}, some of which are discussed in Sec. \ref{ruostekoski_app}.

\section{Methodology}

We describe non-equilibrium quantum dynamics of bosonic atoms by the evolution that can be considered with a good approximation to be unitary. Quantum and thermal noise enter the equations of motion for an atomic field only through the stochastic initial state. The dynamics is governed by the Gross-Pitaevskii Equation (GPE),
\begin{equation}
i\hbar \frac{\partial}{\partial
t}\Phi=\left(-\frac{\hbar^{2} \nabla^2}{2m}
+V_{\rm ext} + g
|\Phi|^{2}\right)\Phi,
\label{ruostekoski_GPE}
\end{equation}
where in 1d the interaction strength $g\rightarrow g_{1d}=2\hbar \omega_{\perp}a$, the $s$-wave scattering length $a$, and $V$ is the trapping potential \cite{olshanii_98}. We concentrate on 1d dynamics in tightly-confined atom traps. In higher dimensions, the unitary evolution may typically be replaced by a model with an explicit low-momentum cut-off \cite{sinatra_lobo_02}. Other realizations of TWA have, e.g., treated the system and environment separately with a coupling between the two, that results in an explicit dynamical noise term in each time step \cite{drummond_hardman_93}, or by including a continuous quantum measurement process \cite{javanainen_ruostekoski_11}. Here, depending on the physical problem, we could also include additional terms in Eq.~(\ref{ruostekoski_GPE}), e.g., atom losses via collisions and spontaneous emission that would generate  also dynamical noise terms for each time step.
In Eq.~(\ref{ruostekoski_GPE}) we have explicitly included the atom number in the nonlinear coefficient, so we use the normalization $\int dx\, |\Phi(x)|^2 = N +m/2$, where $m$ denotes the number of modes in the initial state and $N$ the total number of atoms.

Unlike in the usual GPE, here $\Phi(x,t)$ should be considered as a stochastic phase-space representation of the full field operator describing the time evolution of the ensemble of Wigner distributed wavefunctions. The time evolution is unraveled into stochastic trajectories, where the initial state of each realization
for the classical field $\Phi$ is stochastically sampled in order to synthesize the quantum statistical correlation functions for the initial state.

Since for the unitary evolution all the noise is incorporated in the initial state, it is especially important that the quantum mechanical correlation functions for the initial state of the atomic field operator are synthesized as accurately as practical for each particular physical problem. Here we follow our basic formalism of Refs.~\cite{isella_ruostekoski_05,isella_ruostekoski_06,shrestha_javanainen_09,martin_ruostekoski_10a,martin_ruostekoski_10b,gross_esteve_11}.

\subsection{Initial State Generation in the TWA}

\subsubsection{Uniform system}

In the case of a weakly interacting bosonic gas in a uniform space at $T=0$ the simplest approach to model quantum fluctuations of the atoms, if we are not interested in the conservation of the total atom number, is
the Bogoliubov approximation. In the Bogoliubov theory we calculate the linearized fluctuations of the ground state (or a stationary GPE solution) in which case the back-action of the excited-state atoms on the ground state is ignored \cite{pethick_smith_book_02}. We write the decomposition
\begin{equation}
\hat{\Psi}(x)=\phi_0(x) \hat b_0 +\hat{\psi}'(x)\,,
\label{ruostekoski_V11}
\end{equation}
where the total number of ground state atoms $
N_c=\langle\hat b_0^\dagger \hat b_0\rangle$. The fluctuation part $\hat{\psi}'$ for the excited states can be written in terms of quasi-particle  operators $\hat{b}_q$ and $\hat{b}_q^\dagger$ as
\begin{equation}
\hat{\psi}'(x,t)=\frac{1}{\sqrt L}\,\sum_{q\neq0} (u_q\hat{b}_q \text e^{i q x }-v_q^*\hat{b}_q^\dagger\text e^{-i qx })\,.
\label{ruostekoski_V12}
\end{equation}
The normal mode frequencies $\epsilon_q$ and the quasi-particle amplitudes $u_q$ and $v_q$ can be solved straightforwardly \cite{pethick_smith_book_02} and
the
number of excited-state atoms in the Bogoliubov theory is
\begin{equation}
N'=\sum_{q}\left(|u_q|^2+|v_q|^2\right) n_{\rm BE}(\epsilon_q)+\sum_{q}|v_q|^2,
\label{ruostekoski_V13}
\end{equation}
with
$
\langle \hat b_q^\dagger \hat b_q \rangle=n_{\rm BE}(\epsilon_q)=[\exp{(
\epsilon_q/k_BT)}-1]^{-1}
$ denoting the ideal Bose-Einstein distribution. At $T=0$ we have $n_{\rm BE}(\epsilon_q)=0$.

In order to construct the initial state for the atoms in the TWA evolution
we replace the quantum field operators $(\hat{\Psi},\hat{\Psi}^{\dagger})$ by the classical fields $(\Phi,\Phi^{*})$ by using complex stochastic variables $(\beta_q,\beta_{q'}^*)$ in the place of the quantum operators $(\hat{b}_q,\hat{b}_{q'}^\dagger) $ in Eq.~(\ref{ruostekoski_V12}).

In the Bogoliubov theory, the operators $(\hat{b}_q,\hat{b}_q^\dagger)$ form a set of
ideal harmonic oscillators and
at $T=0$, $(\beta_q,\beta_{q}^*)$ (for $q\neq 0$) are obtained by sampling the corresponding Wigner distribution function \cite{gardiner_zoller_book_99}
\begin{equation}
W(\beta_q,\beta_q^*)=\frac{2}{\pi}\exp[-2|\beta_q|^2]\,.
\label{ruostekoski_V16}
\end{equation}
In this case each unoccupied excitation mode is uncorrelated with Gaussian-distributed noise. The expectation value $\langle \beta_q^* \beta_q\rangle_e=1/2$ specifies the width of the distribution and represents vacuum noise, resulting from the symmetric ordering of the expectation values in the Wigner representation. The noise is distributed in space according to the plane waves,
with a constant density. In the absence of any correlations between the modes, the vacuum noise in the uniform space could be replaced by uncorrelated Gaussian noise on evenly-spaced numerical grid points. However, if we do not want to allow the total atom number to fluctuate between different trajectories (conserved atom number), the simplest modification to the Bogoliubov expansion is to fix the total atom number in each stochastic realization. This introduces long-wavelength correlations between the ground-state mode and the excited-state phonon modes, so that already in this simple example there exist non-trivial spatial noise correlations \cite{martin_ruostekoski_10a,martin_ruostekoski_10b}. For each stochastic realization the number of excited-state atoms satisfies
\begin{equation}
N_s'=\sum_{q}\left(|u_q|^2+|v_q|^2\right) \left(\beta_q^* \beta_q-\frac{1}{2}\right)+\sum_{q}|v_q|^2.
\label{ruostekoski_V17}
\end{equation}
where $N_s'$ fluctuates in each realization with the ensemble average $\langle N_s' \rangle_e=N'=\sum_{q}|v_q|^2$ at $T=0$.
In Eq.~(\ref{ruostekoski_V17}) we have transformed the symmetric ordering of the Wigner representation to quantum expectation values of normally-ordered operators by subtracting $\langle \beta_q^* \beta_q\rangle_e=1/2$ from each mode. The ground-state atom number is then obtained from the fixed total atom number $N$, so that in each stochastic realization $N_{c}=N-N_s'$ and we set $\beta_{0}=\sqrt{ N_c +1/2}$. The ensemble average of the ground-state population is obtained from $ \langle N_c \rangle_e =N- \langle N_s' \rangle_e=N-N'$.

At $T\neq0$ we replace Eq.~(\ref{ruostekoski_V16}) by \cite{gardiner_zoller_book_99}
\begin{equation}
\label{ruostekoski_wigner} W(\beta_q,\beta_q^*)=\frac{2}{\pi}\tanh \left( \epsilon_q /2k_B T\right)
\exp\left[ -2|\beta_q|^2\tanh\left( \epsilon_q /2k_B T\right)\right]\,.
\end{equation}
The Wigner function is Gaussian-distributed with  width ${n}_{\rm BE}(\epsilon_q)+1/2$. The formula introduces thermal populations of each quasi-particle mode and generates more complex spatial noise correlations. After the noise generation, the initial state for stochastic evolution at time $t=0$ may be written as
\begin{equation}
{\Phi} (x)=\phi_0(x) \beta_0 +\frac{1}{\sqrt{L}}\sum_{q\neq0} (u_q {\beta_q} e^{iqx }-v_q^*{\beta_q}^* e^{-i qx})\,.
\end{equation}
Here ${\Phi} (x)$ is a stochastic representation of the full field operator for the atoms.

\subsubsection{Trapped Gases}

Placing atoms in a non-uniform potential results in a spatially-varying initial noise distribution even at $T=0$. For a combined harmonic trap and optical lattice we write the
external potential as $V(x)=m\omega^2 x^2/2 + sE_R \sin^{2}(\pi x/d)$,
where $E_{R}=\hbar^{2}\pi^{2}/2md^{2}$ is the lattice photon recoil energy and $d$ is the lattice period. The Bogoliubov equations now become spatially dependent and need to be solved numerically \cite{pethick_smith_book_02}.
In Eq.~(\ref{ruostekoski_V12}) we replace $u_q e^{i q x}/\sqrt{L}\rightarrow u_j(x)$ and $v_q e^{i q x}/\sqrt{L}\rightarrow v_j(x)$, where the index $j$ refers to the mode number.
In the lowest order approximation the quasi-particle
mode functions $u_j(x)$ and $v_j(x)$ are obtained in the Bogoliubov theory. In several cases of interest where the multi-mode structure of the excitations become important, the Bogoliubov approximation is insufficient due to the large contribution of the
quadratic fluctuation terms.  One consequently needs to use a higher-order theory
in which case the ground-state
and the excited-state populations are solved self-consistently.
One such candidate is the
gapless Hartree-Fock-Bogoliubov (HFB) formalism:
this is similar to the usual HFB approach, but constructed in such a manner that there is no gap in its excitation spectrum at zero momentum.
 The coupled equations for
the ground state and excitations thus take the general form \cite{hutchinson_dodd_98,proukakis_morgan_98}
\begin{eqnarray}
&& \left(\hat{\mathcal{L}}-U_{\mbox{\scriptsize c}}\bar{N}_c|\phi_0|^{2}\right)\phi_0=0\label{ruostekoski_Eqn_GPE}\\
\label{ruostekoski_Bogo} && \hat{\mathcal{L}}
u_j-U_{\mbox{\scriptsize c}}\bar{N}_c\phi_0^2 v_j
 = \epsilon_j u_j,\nonumber\\
&&  \hat{\mathcal{L}}
v_j-U_{\mbox{\scriptsize c}}\bar{N}_c\phi_0^{*2} u_j =-\epsilon_j v_j\,.
\end{eqnarray}
where $u_j(x)$ and $v_j(x)$ ($j>0$) are restricted to the subspace orthogonal to $\phi_0$. In order to express these in a form amenable to stochastic simulations, we have used  the notation $N_c \rightarrow \bar{N}_c = \langle N_c \rangle_e$.
Here
\begin{equation}
\hat{\mathcal{L}}\equiv -{\hbar^2\over 2m}{\partial^2\over\partial x^2}+V(x)+2 U_{\mbox{\scriptsize c}}\bar N_{c}|\phi_0|^2+2U_{\mbox{\scriptsize e}}n'(x)-\mu,
\end{equation}
and $\mu$ is the chemical potential.
This general notation contains numerous theories as sub-cases for appropriate choices of the interaction strength of a condensed atom with another condensed atom ($U_c$) or a thermal atom ($U_e$):
the Bogoliubov approximation is obtained by setting $U_{\mbox{\scriptsize c}}=g_{1d}$, $U_{\mbox{\scriptsize e}}=0$ in Eq.~(\ref{ruostekoski_Bogo}).
Setting $U_{\mbox{\scriptsize c}}=g_{1d}\left[1+m'(x)/\bar{N}_c\phi_0^{2}\right]$ and $U_{\mbox{\scriptsize e}}=g_{1d}$ yields the gapless HFB theory (the G1 version in Ref.~\cite{proukakis_morgan_98}); here $n'(x)=\langle \hat{\psi}'^{\dagger}(x)\hat{\psi}'(x)\rangle$ is the depleted density, and $m'(x)=\langle \hat{\psi}'(x)\hat{\psi}'(x)\rangle$ the anomalous pair correlation, both of which introduce back-action of the excitations on the ground-state. Hence, Eqs.~(\ref{ruostekoski_Eqn_GPE}) and (\ref{ruostekoski_Bogo}) must be solved iteratively until the solutions converge.
In the non-uniform case the number of excited-state atoms is given by
\begin{equation}\label{ruostekoski_depletion2}
N'=\int dx \sum_j \left[ \left(|u_{j}(x)|^{2}+|v_{j}(x)|^{2}\right){n}_{\rm BE}(\epsilon_j)+|v_{j}(x)|^{2}\right],
\end{equation}
and the total atom number may be fixed in each realization as in the uniform case \cite{martin_ruostekoski_10a,martin_ruostekoski_10b}. The gapless HFB theory was introduced as a stochastic sampling technique for TWA simulations in Ref.~\cite{gross_esteve_11} to model reduced atom number fluctuations, fragmentation and spin-squeezing in optical lattice systems.

\subsubsection{Quasicondensate Description}

In tightly-confined 1d traps, the phase fluctuations may be enhanced compared to those obtained using the standard Bogoliubov theory \cite{kheruntsyan_gangardt_03}. A more accurate description can be calculated using quasi-condensate formalism \cite{mora_castin_03} that can be particularly important, e.g., to phase kinks \cite{martin_ruostekoski_10a,martin_ruostekoski_10b}. In the quasi-condensate description we write the field operator as
\begin{equation}
\hat{\Psi}(x) =\sqrt{n_{0}(x)+\delta{\hat{n}}(x)}\exp[i\hat{\theta}(x)].\label{ruostekoski_Quasi}
\end{equation}
The density $\delta{\hat{n}}(x)$ and phase $\hat{\theta}(x)$ operators are written as (for $j>0$)
\begin{eqnarray}
\hat{\theta}(x) &=&
-\frac{i}{ 2\sqrt{n_{0}(x)}}
\sum_{j}\left(\theta_{j}(x) \hat{b}_{j}-\theta_{j}^{*}(x)\hat{b}_{j}^{\dagger}\right),\label{ruostekoski_phaseop}\\
\delta\hat{n}(x)  &=& \sqrt{n_{0}(x)}\sum_{j}\left(\delta n_{j}(x)\hat{b}_{j}+ \delta n_{j}^{*}(x)\hat{b}_{j}^{\dagger}\right)\,,\label{ruostekoski_densityop}
\end{eqnarray}
%
where $\theta_{j}(x)=u_{j}(x)+v_{j}(x)$ and $\delta n_{j}(x)=u_{j}(x)-v_{j}(x)$ are given in terms of the solutions to the Bogoliubov equations (see the previous section). This results in a stochastic Wigner representation $(\theta_W(x),\delta n_W(x))$ of phase and density operators. The stochastic initial state at $t=0$ for the time evolution then reads \cite{martin_ruostekoski_10a,martin_ruostekoski_10b}
\begin{equation}
{\Phi}(x) =\sqrt{n_{0,W}(x)+\delta{{n}}_W(x)}\exp(i{\theta}_W(x)),\label{ruostekoski_twainitial}
\end{equation}
where the ground-state density $n_{0,W}(x)=(N_{qc}+1/2)|\phi_0(x)|^2$.

\subsubsection{Relaxation}

We may also consider an ideal, non-interacting BEC as an initial state for the TWA simulations,
but before the actual time evolution, we can continuously turn up the nonlinear interactions between
the atoms. If the process is slow enough and relaxes to the ground state, we may be able to produce
the stochastic initial state of the interacting system. Although this may simplify the calculations, in practice the technique in a closed system does not necessarily converge to the correct interacting state \cite{isella_ruostekoski_06}. More complex models with open systems, kinetic equations and time-dependent noise
can help the relaxation process at finite temperatures \cite{cockburn_nistazakis_10}.

\subsection{Wigner Representation and Symmetric Ordering}

The Wigner distribution returns symmetrically-ordered expectation values of any stochastic representations of quantum operators. In particular, the expectation values of the full multi-mode Wigner fields in the TWA simulations of the time-dynamics are symmetrically ordered with respect to every mode. In general, this can significantly complicate the analysis of the numerical results when quantum fluctuations are important \cite{isella_ruostekoski_06}. A numerically practical transformation of the symmetrically-ordered expectation values to the normally-ordered expectation values of physical observables can be done using projection techniques (see Refs.~\cite{isella_ruostekoski_05,isella_ruostekoski_06,gross_esteve_11}). In the presence of an optical lattice, a natural approach is to project the stochastic field on to the several lowest mode functions of the individual lattice sites. We denote the annihilation operator for the atoms in the $j$th vibrational mode of the site $i$ as $\hat a_{i,j}$. We write the corresponding stochastic amplitude as $a_{i,j}$ which can numerically be obtained from
\begin{equation}
\label{ruostekoski_projection} a_{i,j}(t)=\int_{i^{\rm th} {\rm well}}dx\,
[\varphi_{i,j}(x,t)]^*\Phi(x,t)\,,
\end{equation}
where $\Phi$ is the stochastic field
and $\varphi_{i,j}$ is the $j$th vibrational mode function
of the site $i$.
The integration is performed over the $i$th site and the normally ordered quantum expectation values for the site populations reads
\begin{equation}
\langle\hat n_i\rangle= \sum_j \langle \hat a_{i,j}^\dagger \hat a_{i,j}\rangle=\sum_j \big[ \langle a_{i,j}^*  a_{i,j}\rangle_e-1/2 \big],
\end{equation}
Fluctuations are calculated using analogous transformations. For the on-site fluctuations of the atom number in the $i$th site we obtain
\begin{eqnarray}
( \Delta n_{i})^{2}&=& \langle \hat{n}_{i}^{2} \rangle-\langle \hat{n}_{i} \rangle^2\nonumber\\
 &=& \sum_{j,k}\left[\langle a_{i,j}^{*} a_{i,j} a_{i,k}^{*} a_{i,k} \rangle_{e}-\langle a_{i,j}^{*} a_{i,j}\rangle_{e}\langle a_{i,k}^{*} a_{i,k} \rangle_{e}-\delta_{jk} /4\right].\label{ruostekoski_Eq_deltan2}
\end{eqnarray}
Similarly, the relative atom number fluctuations between the sites $p$ and $q$ are obtained
from
\begin{eqnarray}
  \left[\Delta (\hat{n}_{p}-\hat{n}_{q})\right]^2
&= &\sum_{i,k}\left[\left\langle \left( a_{p,i}^{*}a_{p,i}-a_{q,i}^{*}a_{q,i}\right)
\left(a_{p,k}^{*}a_{p,k}-a_{q,k}^{*}a_{q,k}\right) \right\rangle_{e}\right.\nonumber\\
&-& \left. \langle a_{p,i}^{*}a_{p,i}-a_{q,i}^{*}a_{q,i}\rangle_{e}
\langle a_{p,k}^{*}a_{p,k}-a_{q,k}^{*}a_{q,k} \rangle_{e}-\delta_{ik}/2\right]. \label{ruostekoski_Eq_deltaJ}
\end{eqnarray}
Alternatively, we could have, for instance, written
\begin{eqnarray}
\langle\hat n_j\rangle = \int_{j }dx\, \langle \hat \Psi^\dagger(x)\hat \Psi(x)\rangle &=& \int_{j}dx\, \langle \Phi^*(x) \Phi(x)\rangle_{e} \nonumber \\ &-& {1\over 2} \int_{j }dx\,
\sum_i \left( |u_i(x)|^2 - |v_i(x)|^2\right).
\end{eqnarray}
Calculation of $\langle\hat n_j^2\rangle$ then, however, results in double integrals over the sites
that can be computationally slow when performed over a large number of realizations.

\subsection{Numerical Implementation}

In the numerical implementation the initial state fluctuations are solved by first finding a (stable) stationary state or a local energetic minimum in GPE. If the system is assumed to be initially in thermal equilibrium, we find the corresponding ground state by imaginary time evolution of GPE, e.g., by using the nonlinear split-step Fourier methods \cite{javanainen_ruostekoski_06}. Using the ground state solution we then diagonalize the linearized equations for the quasi-particle excitations in order to obtain the eigenfunctions $u_j(x),v_j(x)$ and
the corresponding eigenenergies. In the case of self-consistent HFB method, the excitations and the ground state are solved iteratively until the solutions converge \cite{hutchinson_dodd_98}.

The time evolution of the ensemble of Wigner distributed wavefunctions is unraveled into stochastic trajectories, where the initial state of each realization for the classical stochastic field $\Phi$ is generated using the quasiparticle mode functions and amplitudes \cite{isella_ruostekoski_06}. The complex, Gaussian-distributed stochastic mode amplitudes are sampled using the Box-Muller algorithm \cite{press_teukolsky_book_02}. During the time evolution we simulate some physical process that describes the changing of the equilibrium configuration, e.g., displacement of atoms from the trap center \cite{ruostekoski_isella_05} or turning up of the optical lattice potential \cite{isella_ruostekoski_05,isella_ruostekoski_06}.
The integration of the time dynamics is also performed using the nonlinear split-step methods \cite{javanainen_ruostekoski_06}, typically on a spatial grid of a few thousand grid points. In several cases the sufficient convergence is obtained after 600-1000 realizations.

In order to transform the symmetrically-ordered expectation values of the Wigner representation into normally-ordered expectation values, we  numerically introduce an orthonormal basis, e.g., in each lattice site. The stochastic field is then at different times projected onto this basis and the desired expectation values are evaluated using the transformations for each projected mode function, as described in the previous section.

\section{Validity Issues}

In 2d and 3d the TWA can have implementation problems.  Firstly,  the
atom cloud can heat during time evolution due to rapid nonlinear dynamics between the vacuum modes \cite{sinatra_lobo_02}.  Secondly,   physical observables can diverge as a function of the number of modes (or equivalently energy cut-off or grid spacing). Importantly for the present discussions, 1d systems are more robust  to these effects.
In particular, TWA has been successful in describing superfluid dynamics in the presence of considerable quantum fluctuations in 1d systems, even though it is
clearly insufficient, e.g., in a Mott-insulator regime
and at very low atom numbers.
For instance, TWA simulations \cite{ruostekoski_isella_05} were qualitatively able to produce the experimentally observed damping rate of center-of-mass oscillations of bosonic atomic cloud in
a very shallow, strongly confined 1D optical lattice, corresponding to the dissipative atom transport experiments of Ref.~\cite{fertig_ohara_05} in which case atom numbers approximately down to 70-80 were used in an elongated trap of a very large phase space.

The accuracy of the initial state noise generation can
be a crucial limitation, especially in simulations involving very short time dynamics. The spatial distribution of phonon excitations in trapped systems can result in very rapid noise variation where, e.g., phase fluctuations dominate near the edges of the atom cloud \cite{isella_ruostekoski_06}. For dark soliton dynamics, the differences in the soliton trajectories between the cases in which the noise was generated within the quasi-condensate description and in the Bogoliubov theory are notable \cite{martin_ruostekoski_10a}, indicating that the soliton imprinting process and dynamics are sensitive to enhanced phase fluctuations of the quasi-condensate description \cite{mora_castin_03}. Evaluating phonon modes in the linearized Bogoliubov approximation may also become inaccurate, compared to self-consistent HFB methods even at $T=0$, as demonstrated in the case coupled condensates in a few-site lattice system \cite{gross_esteve_11}.

Stochastic phase-space methods based on sampling classical probability distributions are necessarily approximate, unless the problem is reformulated, e.g., by doubling the phase space and considering $\Phi$ and $\Phi^*$ as independent fields. Such positive-P \cite{drummond_gardiner_80,gardiner_zoller_book_99,carusotto_castin_01,drummond_deuar_04} or positive-Wigner \cite{plimak_olsen_01} methods can in principle provide exact solutions but frequently run into numerical problems due to rapidly growing sampling errors.

\section{Applications \label{ruostekoski_app}}

\subsection{Dark Solitons \label{ruostekoski_solitons}}

Dark solitons have been actively studied in BECs \cite{burger_bongs_99,denschlag_simsarian_00,anderson_haljan_01,dutton_bude_01,weller_ronzheimer_08,stellmer_becker_08},  and in nonlinear optics \cite{kivshar_luther-davis_98}. Although there exist numerous studies of classical solitons, the quantum properties of dark solitons are much less known. Numerical TWA simulations are suitable for the studies of the creation and non-equilibrium quantum dynamics of solitons in 1d traps.
We consider the experimental imprinting method \cite{burger_bongs_99,denschlag_simsarian_00}, where a soliton is generated by applying a `light-sheet potential', of value $V_{\phi}$ to half of the atom cloud, for time $\tau$, so that in the corresponding classical case the light sheet imprints a phase jump of $\phi_c=V_{\phi}\tau/\hbar$ at $x=0$, preparing a dark soliton. Classically the imprinted soliton oscillates in a harmonic trap at the frequency $\omega/\sqrt{2}$ \cite{busch_anglin_00} with
the initial velocity ${\bf v}/c=\cos(\phi_c/2)$, depending on $\phi_c$ and the speed of sound $c$. The soliton is stationary (dark) for $\phi_c=\pi$, with  zero density at the kink. Other phase jumps produce moving (grey) solitons, with non-vanishing densities at the phase kink.

In TWA simulations we generate the initial state using the quasi-condensate formalism and vary the ground-state depletion $N' /N$. At $T=0$ we keep the nonlinearity $Ng_{1d}$ fixed, but adjust the ratio $g_{1d}/N$. This is tantamount to varying the effective interaction strength $\gamma_{\rm int}=m g_{1d}/\hbar^2 n$ \cite{kheruntsyan_gangardt_03}. We can also study the effects of thermal depletion by varying $T$.

\begin{figure}
\centerline{\includegraphics[width=10.0cm]{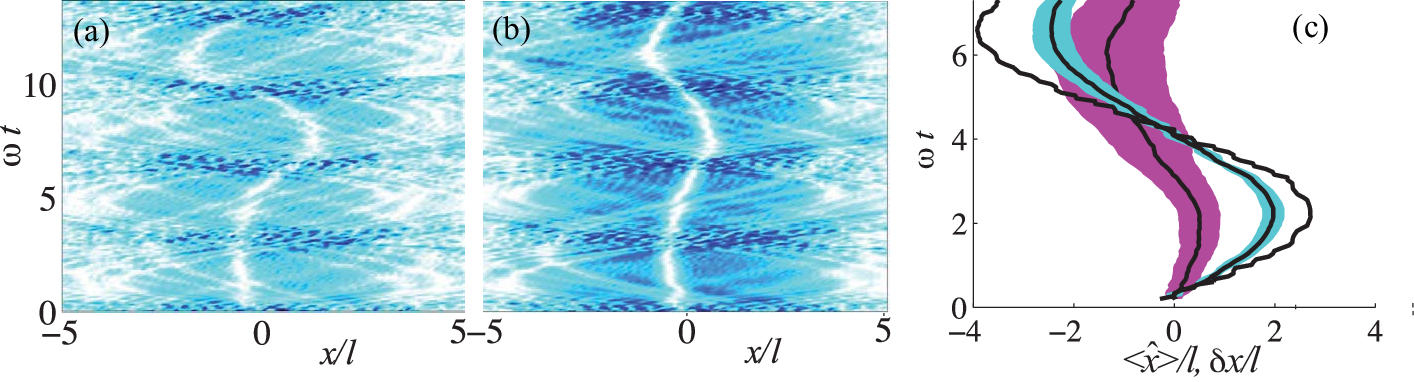}}
\caption{Soliton dynamics in a harmonic trap showing
(a-b) the Wigner density $|\Phi(x,t)|^{2}$ for individual stochastic realizations with the same $g_{1d}N=100\hbar\omega l$, $\phi_{c}=2$, and $T=0$ for $N=50,100$, in (a) and (b), respectively; (c) The quantum mechanical expectation value for the soliton position $\langle \hat x\rangle$ (solid lines) and its uncertainty $\Delta x$ (shaded regions) for $N=8000,440,50$ (curves with decreasing amplitudes) with the same nonlinearity $g_{1d}N$, $\phi_{c}=2$, and $T=0$. At $N\simeq8000$, $\Delta x$ is negligible. Quantum fluctuations increase $\Delta x$ and soliton damping, and decrease the speed. Figure taken from Ref.~\cite{martin_ruostekoski_10a}
\label{ruostekoski_fig1} }
\end{figure}

In the presence of noise, soliton trajectories in the TWA fluctuate between different realizations due to quantum and thermal fluctuations \cite{dziarmaga_karkuszewski_03,mishmash_carr_09,martin_ruostekoski_10a,martin_ruostekoski_10b,cockburn_nistazakis_10}. Individual stochastic realizations of $|\Phi|^2$ in a harmonic trap represent possible experimental observations of single runs \cite{isella_ruostekoski_06,martin_ruostekoski_10a,martin_ruostekoski_10b}. In the TWA we can ensemble average hundreds of stochastic realizations in order to obtain quantum statistical correlations of the soliton dynamics. We numerically track the position of the kink at different times in individual realizations and calculate the quantum mechanical expectation values for the soliton position $\langle \hat x \rangle$ and its uncertainty $\Delta x=\sqrt{\langle \hat x^{2}\rangle-\langle \hat x\rangle^2}$ \cite{martin_ruostekoski_10a,martin_ruostekoski_10b}.
Our results are summarized in Fig. \ref{ruostekoski_fig1}.

\subsection{Atom Number Squeezing \label{ruostekoski_squeezing}}

\begin{figure}
\centerline{\includegraphics[width=10.0cm]{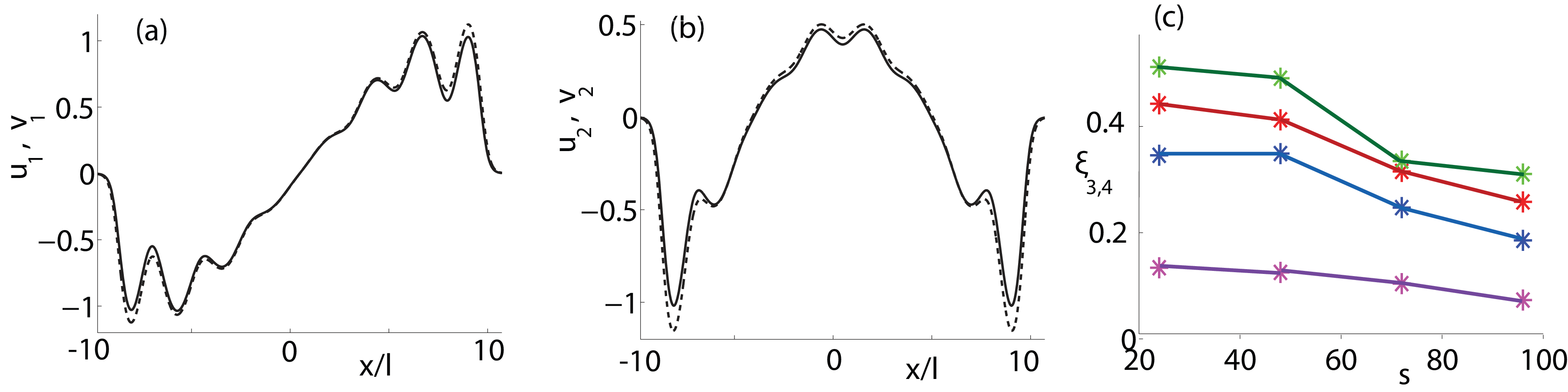}}
\caption{The numerical solution of the lowest two HFB modes in a six-site optical lattice showing
(a) $u_1(x)$ (dotted) and $v_1(x)$ (solid); (b) $u_2(x)$ (dotted) and $v_2(x)$ (solid) at $s=24$ and $T\simeq5.5$nK; (c) Relative atom number (or spin) squeezing at different lattice height between two central nearest-neighbor sites $\xi_{pq}=\left[\Delta (\hat{n}_{p}-\hat{n}_{q})\right]^2 (n_p+n_q)/(4n_pn_q)$.
The different data sets correspond to temperatures (curves from top to bottom) $T\simeq5.5$nK, $T\simeq4.5$nK, $T\simeq4.0$nK, and $T=0$. The harmonic trap frequency is
$\omega=2\pi\times21$Hz, the atom number $N\simeq 5000$, and the lattice spacing $d\simeq 5.7\mu$m.}
\label{ruostekoski_fig2}
\end{figure}

In this section we consider an example of a TWA calculation of spin and relative atom number squeezing due to turning up of an optical lattice. Unlike in
Refs.~\cite{isella_ruostekoski_05,isella_ruostekoski_06} where a BEC fragmentation in TWA was investigated by a lattice with a large number of small sites, we simulate a six-site system, analogous to the recent experimental observations of spin and relative atom number squeezing \cite{esteve_gross_08,gross_esteve_11} as well as reduced on-site atom number fluctuations and long-range correlations \cite{gross_esteve_11} between coupled condensates. Bose-condensed $^{87}$Rb atoms are confined to a cigar-shaped optical dipole trap where an optical lattice is applied along the axial direction. The lattice potential is slowly turned up from $s(0)=48 E_R$ to $s(\tau)=96E_{R}$. 
Due to large individual lattice sites the multi-mode structure of the fluctuations is important, and the atom number fluctuations are evaluated by using the projection technique on to several modes in each site, as explained in the previous section. The Bogoliubov approximation is not accurate due to phonon-phonon interactions, indicating a significant contribution of higher-order terms to atom number fluctuations even at $T=0$, and the initial state is calculated using the HFB method \cite{gross_esteve_11}. The TWA simulation results demonstrated a qualitative agreement with the experimental observations, although the experiment was not performed in a tightly-confined 1d trap with completely suppressed radial density oscillations. The spatially non-uniform distribution of quantum and thermal fluctuations is clearly seen in Fig.~\ref{ruostekoski_fig2}. The lowest HFB modes dominantly occupy the outer regions of the atom cloud with significantly enhanced atom number and phase fluctuations in those sites. Such fluctuations could not be represented, e.g., by a uniform stochastic noise sampling.

\section{Comparisons}

\begin{figure}
\centerline{\includegraphics[width=10.0cm]{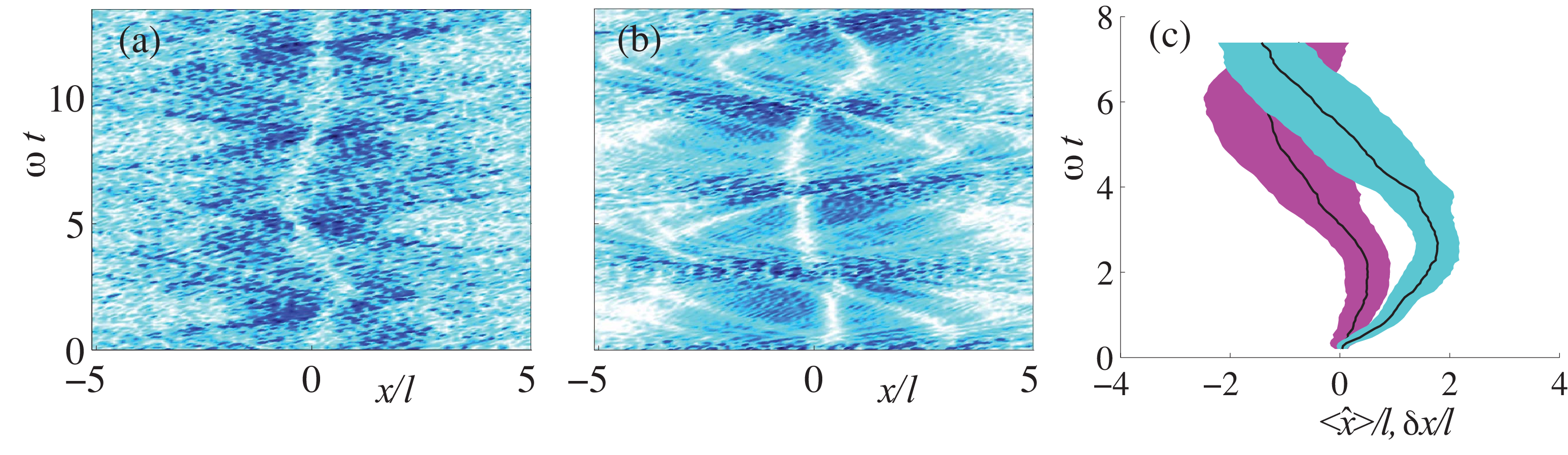}}
\caption{Differences between initial noise generation in TWA. Dark soliton dynamics in a 1d harmonic trap showing
(a-b) the Wigner density $|\Psi_{w}(x,t)|^{2}$ for individual stochastic realizations of TWA for  $\phi_{c}=2$, $T=0$, and $N=50$ (parameters as explained in \ref{ruostekoski_solitons}). In (a) the initial state is generated within the Bogoliubov approximation and in (b) using the quasi-condensate formalism; (c) The quantum mechanical expectation values for the soliton position $\langle \hat x\rangle$ (solid lines) and its uncertainty $\delta x$ (shaded regions). The lighter curve with larger oscillation amplitude corresponds to the Bogoliubov case and the darker one the quasi-condensate case.}
\label{ruostekoski_extrafig1}
\end{figure}

The accuracy of the initial state noise generation can contribute to the simulation results. For dark soliton dynamics, the differences in the soliton trajectories between the cases in which the noise was generated within the quasi-condensate description and in the Bogoliubov theory are notable \cite{martin_ruostekoski_10a}. Some examples are illustrated in Fig.~\ref{ruostekoski_extrafig1}. As previously noted, also evaluating phonon modes in the linearized Bogoliubov approximation may become inaccurate, compared to self-consistent HFB methods.

\section*{Acknowledgments}

We acknowledge financial support from EPSRC and Leverhulme Trust.

\bibliographystyle{apsrev4-1}

\bibliography{ruostekoski}

\end{document}